\documentclass[twocolumn,showpacs,aps,prl,superscriptaddress]{revtex4}

\usepackage{graphicx}
\usepackage{dcolumn}
\usepackage{amsmath}
\usepackage{epsfig}
\usepackage{ulem}

\input pubboard/babarsym

\long\def\inst#1{\par\nobreak\kern 4pt\nobreak
    {\it #1}\par\vskip 10pt plus 3pt minus 3pt}

\def\Lc     {\ensuremath{\Lambda_c^+}\xspace}
\def\Lcb    {\ensuremath{\kern  0.18em 
               \overline{\kern -0.18em \Lambda}\kern 0.05em _c^-}\xspace}
\def\Lccb   {\ensuremath{^{(}\kern  0.18em 
               \overline{\kern -0.18em \Lambda}\kern 0.05em _c^-}\xspace}
\def\LcLcb  {\ensuremath{\Lc\Lcb X}\xspace}
\def\pKpi   {\ensuremath{p\Km\pip}\xspace}
\def\pKsh   {\ensuremath{p\KS}\xspace}
\def\LcpKpi {\ensuremath{\Lc \!\!\to\! p\Km\pip}\xspace}
\def\LcpKsh {\ensuremath{\Lc \!\!\to\! p\KS}\xspace}

\def\Xc0    {\ensuremath{\Xi_c^0}\xspace}
\def\pstar  {\ensuremath {p^*}\xspace}

\def\eeqqb  {\ensuremath{\epem \!\!\to \! q\overline{q}}\xspace}
\def\eeccb  {\ensuremath{\epem \!\!\to \! c\overline{c}}\xspace}
\def\pbar   {\ensuremath{\overline{p}}\xspace}
\def\ppbar  {\ensuremath{p/\overline{p}}\xspace}

\def\drap   {\ensuremath{| \Delta y |}\xspace}
\def\bary   {\ensuremath{N}\xspace}
\def\aary   {\kern 0.18em\overline{\kern -0.18em N}{}\xspace}

\newcommand{\gevcsq}{\ensuremath{{\mathrm{\,Ge\kern -0.1em V^2\!/}c^4}}\xspace}

\newcommand{\BABARPubYear}    {07}
\newcommand{\BABARPubNumber}  {027}

\newcommand{\SLACPubNumber} {14157}

\begin{document}

\preprint{\babar-PUB-\BABARPubYear/\BABARPubNumber} 
\preprint{SLAC-PUB-\SLACPubNumber} 

\begin{flushleft}
SLAC-PUB-\SLACPubNumber \\
\babar-PUB-\BABARPubYear/\BABARPubNumber\\
\end{flushleft}


\title{\boldmath
Correlated Leading Baryon-Antibaryon Production
in $e^+e^- \!\!\rightarrow\! c\bar{c} \!\rightarrow\! \Lc \Lcb X$ 
} 

%
\author{B.~Aubert}
\author{Y.~Karyotakis}
\author{J.~P.~Lees}
\author{V.~Poireau}
\author{E.~Prencipe}
\author{X.~Prudent}
\author{V.~Tisserand}
\affiliation{Laboratoire d'Annecy-le-Vieux de Physique des Particules (LAPP), Universit\'e de Savoie, CNRS/IN2P3,  F-74941 Annecy-Le-Vieux, France}
\author{J.~Garra~Tico}
\author{E.~Grauges}
\affiliation{Universitat de Barcelona, Facultat de Fisica, Departament ECM, E-08028 Barcelona, Spain }
\author{M.~Martinelli$^{ab}$}
\author{A.~Palano$^{ab}$ }
\author{M.~Pappagallo$^{ab}$ }
\affiliation{INFN Sezione di Bari$^{a}$; Dipartimento di Fisica, Universit\`a di Bari$^{b}$, I-70126 Bari, Italy }
% ...
\collaboration{The \babar\ Collaboration}
\noaffiliation


\begin{abstract}

We present a study of
$649 \pm 35$
\eeccb events produced at 
$\sqrt{s}\! \approx\! 10.6\, \gev$
containing both a \Lc baryon and a \Lcb antibaryon.
The number observed is roughly four times that expected if the leading charmed hadron 
types are uncorrelated,
confirming an observation by the CLEO Collaboration.
We find a 2-jet topology in these events but very few additional baryons,
demonstrating that the primary $c$ and \cbar are predominantly contained in a
correlated baryon-antibaryon system.
In addition to the charmed baryons
we observe on average $2.6 \pm 0.2$ charged intermediate mesons,
predominantly pions,
carrying 65\% of the remaining energy.

\end{abstract}


\maketitle

Baryon production in high-energy jets from \epem annihilations has 
presented a series of challenges to our understanding of strong interactions.
Its observation led to the competing notions of `primary' and `local' 
baryon correlations~\cite{oddone}.
In the former, 
the $e^+$ and $e^-$ annihilate into a primary diquark-antidiquark,
rather than a quark-antiquark,
pair.
The diquark and antidiquark then hadronize into jets containing a
leading baryon $\bary_1$ and a leading antibaryon $\aary_2$,
respectively, but no other (anti)baryons.
$\bary_1$ and $\aary_2$ would then share two quark flavors and typically
have high, antiparallel momenta and large values of 
variables characterizing their separation,
such as invariant mass or rapidity difference \drap, where 
$y \equiv 0.5 \ln[(E + p_\parallel) /(E - p_\parallel)]$,
$E$ is the baryon energy, 
and $p_\parallel$ is the projection of its momentum on the thrust axis.
Alternatively, 
an $\bary_1\aary_2$ pair might be produced locally, 
in an individual step of a hadronization cascade, 
with a smaller value of \drap.
Most experimental studies of baryon-antibaryon pairs have shown \drap 
distributions that peak at small values~\cite{corrlrefs}.

Several mechanisms to describe baryon production and correlations
have been implemented in Monte Carlo hadronization models~\cite{params}.
In the JETSET~\cite{jetset} color-flux-tube model, 
a tube break can result in a diquark-antidiquark (rather than \qqbar)
pair, producing an $\bary_1\aary_2$ pair locally.
  An intermediate meson is introduced between $\bary_1$ and $\aary_2$ 
  with some probability (50\% by default~\cite{lundbook})
  to match the measured \drap distributions.
In the HERWIG~\cite{herwig} model, an individual, color-singlet cluster may
fragment into a baryon-antibaryon pair but not a multi-body state with
additional mesons. The model does not reproduce the measured \drap
distributions when tuned to other observables~\cite{corrlrefs}.
The UCLA~\cite{ucla} area-law model includes $\bary_1\aary_2$ pairs
with any number of intermediate mesons, 
and suppresses higher-mass intermediate meson systems by means of a 
tunable parameter.

Direct evidence of primary production and/or intermediate mesons would be
of great interest, 
but previous searches for the latter using three-particle 
correlations~\cite{delphillk} 
or baryon flavor correlations~\cite{alephxxb} were generally inconclusive.

At center-of-mass (c.m.)\ energies $\sqrt{s}$ much larger than four 
baryon masses, the assumption of
local baryon number conservation implies that
an \eeqqb event containing a leading baryon $\bary_1$ in the $q$ jet 
and a leading antibaryon $\aary_2$ in the \qbar jet 
must also contain an antibaryon $\aary_3$ in the $q$ jet 
and a baryon $\bary_4$ in the \qbar jet.
However, if the $\bary_1\aary_3\bary_4\aary_2$ mass is a large
fraction of $\sqrt{s}$, 
these four-baryon events would be suppressed 
and other processes might be visible---in particular,
  primary baryon production events with exactly two baryons, one in
  each jet.
At $\sqrt{s}\!\approx\! 10$~\gev,
charmed ($c$) baryons are of particular interest,
since any high-momentum $c$ or \cbar baryon must be a leading particle
in an \eeccb event, 
and any $\bary_{c1}\aary_3\bary_4\aary_{c2}$ mass 
exceeds 6.5~\gevcc.
The CLEO Collaboration reported
  an excess by a factor of $3.5 \pm 0.6$~\cite{cleolclcb}
  in the number
of events at $\sqrt{s}\! =\! 10.6$~\gev
with both a \Lc and a \Lcb,
where their expectation is derived
assuming local baryon number conservation in the JETSET model and
from observed events with a \Lc and a $D^-$ or $\overline{D}^0$ meson.
This excess is evidence that the baryon
production is correlated between the $c$ and \cbar jets
and is consistent with primary baryon production, but does not
exclude the possibility of local baryon production with 
  correlation
between the jets. The two cases can be distinguished experimentally:
local production would require an additional baryon and antibaryon
($\bary_4$ and $\aary_3$) in the event, so events with exactly
one \Lc, exactly one \Lcb, and no additional baryons would imply
primary production. CLEO investigated this and did not observe a strong
signal for additional protons in the $\Lc\Lcb$ candidate events,
but due to a limited data sample and the lack of a limit on
additional neutrons they were unable to exclude
local baryon production.

In this paper
we exploit the particle identification capabilities of the 
\babar\ detector~\cite{babarNIM} to select a sample of \LcLcb events 
in which the \Lc and \Lcb are produced at high momentum in opposite
hemispheres, and study their characteristics in detail.
We use 220~\invfb of data collected at $\sqrt{s}\!=\,$10.54--10.58~\gev.
We identify the charged tracks in the $X$ system,
looking for additional (anti)protons,
and search for higher-mass baryons that could be a source
of the \LcLcb events.
We consider charged tracks measured in the silicon vertex tracker (SVT) 
and drift chamber (DCH),
and identified as pions, kaons or protons using the DCH and the 
detector of internally reflected Cherenkov light.
The 
  identification
algorithm used here~\cite{bbrinclLc,brandon} is over 99\%
efficient for pions and kaons (protons) within the acceptance with
momenta between 0.15 and 0.5 (1.2)~\gevc, 
with misidentification rates below 0.5\%.
At higher momenta it remains over 90\%
  efficient,
with misidentification rates generally below 1\%.

We construct \Lc candidates in the $pK^-\pi^+$ and $p\KS$ decay modes
and \Lcb in the corresponding charge-conjugate modes.
We consider a pair of oppositely charged tracks as a 
$\KS \!\!\to\! \pip\pim$ candidate if 
a vertex fit returns a $\chi^2$ with a confidence level (CL) exceeding 0.01,
the vertex is displaced by 2.5--60~cm from the interaction point (IP)
calculated for each event from the set of well-measured tracks in the SVT, 
the angle $\theta_{K_S}$ between
  the \KS candidate's
momentum and the
IP-to-vertex direction satisfies $\cos\theta_{K_S}\! >\! 0.97$,
and the $\pip\pim$ invariant mass is in the range 491.8--503.8~\mevcc.
All combinations of a \KS and a well-measured 
($\geq$15 hits in the DCH and $\geq$5 in the SVT)
proton are considered \LcpKsh candidates.
A combination of well-measured $p$, \Km and \pip tracks is 
considered a \LcpKpi candidate if its vertex fit yields CL$>$0.001.

\begin{figure}[t]
\begin{center} 
\includegraphics[width=\linewidth]{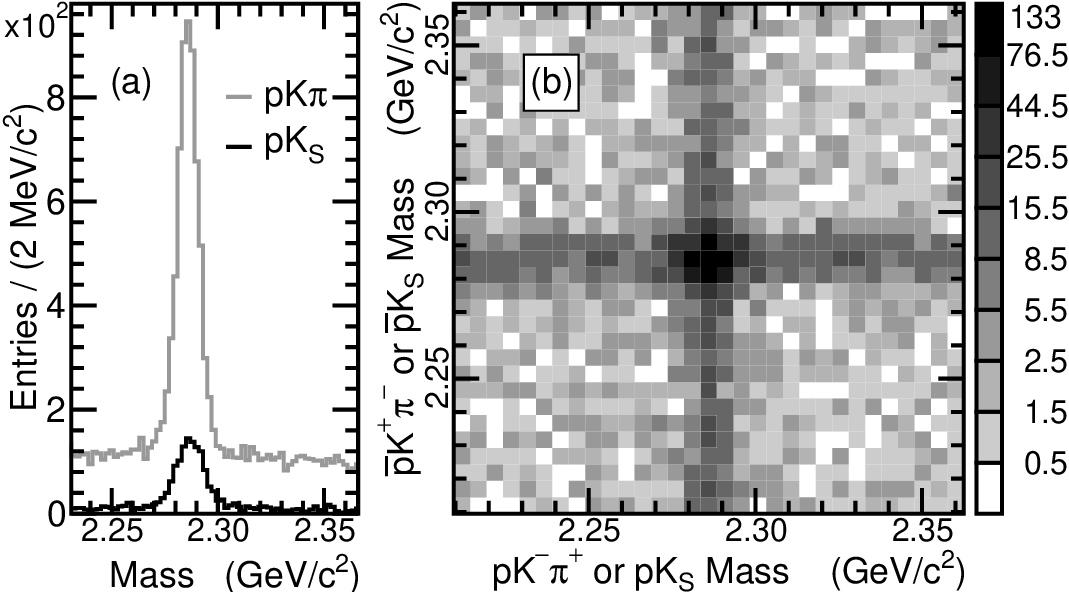}
\end{center}
\vspace{-0.6cm}
\caption {
  (a)
  Invariant mass distributions for the \Lc/\Lcb candidates in 
  selected events, reconstructed in the $pK\pi$ (gray) and
  $p\KS$ (black) decay modes.
  (b)
  Invariant mass of the \Lcb candidate vs.\ that of the 
  \Lc candidate in the same event, in 5~\mevcc square bins.
  }
\label{fig:Lcmass}
\end{figure}

We require \pstar, the momentum of the \Lc candidate in the 
\epem c.m.\ frame, to exceed 2.3~\gevc,
so that the rate of \Lc from \Y4S decays~\cite{bbrinclLc,bellechsp} is negligible.
We select events containing at least one \Lc candidate and at least one \Lcb candidate,
requiring each candidate to have mass within 190~\mevcc of the fitted \Lc peak. We then form
\Lc\Lcb pairs provided that they have no common tracks in their decay chains.
For these 21,000 pairs we show the candidate \pKpi and \pKsh invariant
mass distributions in Fig.~\ref{fig:Lcmass}a.
Clear \Lc signals are visible over modest backgrounds.
The peak mass values, rates, and momentum distributions
are consistent with previous measurements~\cite{bbrinclLc,bellechsp,lcmass}.
We plot the invariant mass of the \Lcb candidate versus that of the \Lc 
candidate in Fig.~\ref{fig:Lcmass}b.
Horizontal and vertical bands are visible, 
corresponding to events with a real \Lcb or \Lc, respectively,
and there is a substantial enhancement where they overlap.

The opening angle $\theta$ between the \Lc and \Lcb momenta in the 
c.m.\ frame is sensitive to their production mechanism.
We expect \Lc\Lcb pairs from gluon splitting 
($\eeqqb g\!\to\! \qqbar\ccbar$)
or $\eeccb g$ events with a very hard gluon
to have relatively small $\theta$,
but also a suppressed selection efficiency due to the $p^*$ requirement.
In the 21,000 events selected, $\theta$ values are concentrated
near 180$^\circ$,
consistent with dominance of 2-jet \eeccb events.
Only 
  seven events have $\theta\! <\! 90^\circ$,
  one 
of which is in the signal region defined below.
Since the small-$\theta$ background may have different characteristics 
from that at large $\theta$, 
we require $\theta\! >\! 90^\circ$.

About 3\% of the events have two \Lc (or two \Lcb) candidates,
due to the two \pKpi combinations in 
the decay chains
$\Sigma_c^{++}  \! \to \! \Lc(\pKpi)\pip$ and
$\Lambda_c^{*+} \! \to \! \Lc(\pKpi)\pip\pim$.
We include all combinations in the sample and account
for the kinematic overlap through the background subtraction.
We define a circular \LcLcb signal region centered at our peak mass values
with a radius of 12~\mevcc,
which contains 919 entries.
Using the single-\Lc/\Lcb bands~\cite{brandon}, 
we estimate an expected background
in the signal region of $245 \pm 5$ events 
with one real \Lc or \Lcb and one fake.
Using events with both masses at least 40~\mevcc from the fitted \Lc mass,
we estimate $25 \pm 1$
  expected
background events with fake \Lc and \Lcb,
giving a \LcLcb signal of
  $N_{\Lc\Lcb} = 649 \pm 35$
events. 

We can calculate an expected number of signal events, $n_{exp}$, under the assumption
that the $c$ and \cbar hadron types are uncorrelated
so that all signal events are four-baryon events.
Then $n_{exp}\! =\! C n_1^2 / 4N_{\ccbar}$,
where 
$n_1\! =\! 420,000$ is the number of
  single
\Lc/\Lcb observed in the data,
$N_{\ccbar}\! =\,$3$\times$10$^8$ is the number of \eeccb events
expected for our integrated luminosity,
and the factor $C$ accounts for the correlation between the \Lc and \Lcb
reconstruction efficiencies.
This formulation is independent of the \Lc branching fractions and
average efficiencies.
In the simple case where the efficiencies of the \Lc and \Lcb in
\LcLcb events are uncorrelated, no correction is needed ($C=1$)
and $n_{exp} = n_1^2 / 4N_{\ccbar}$.
More generally, $0<C<1/\varepsilon$
for an average acceptance times efficiency of $\varepsilon$:
in the extreme case of maximal correlation $C=1/\varepsilon$,
and in the extreme case of maximal anticorrelation $n_{exp}\! = C \! = 0$. 
At \babar\ there might be correlations
because of the asymmetric beam energies and detector
layout. We evaluate this correction using
the JETSET, HERWIG, and UCLA models, adjusting
their charm fragmentation parameters and reweighting the resulting \pstar
distributions to reproduce our measured 
distribution for inclusive \Lc~\cite{bbrinclLc}.
  Combined
with smooth parametrizations of our efficiencies as functions
of momentum and polar angle,
the models
give values of $C$ ranging from 0.63 to 1.65, with a mean of
  $1.05$.
  Even allowing for the large model dependence,
the full range of 
  $n_{exp}\! =\,$100--250
  events is 
  well
below the observed 
  $649 \pm 35$,
confirming the enhanced rate 
$N_{\Lc\Lcb}/n_{exp} \approx 4$
reported by the CLEO Collaboration~\cite{cleolclcb}.

\begin{figure}[t]
\begin{center} 
\includegraphics[width=\linewidth]{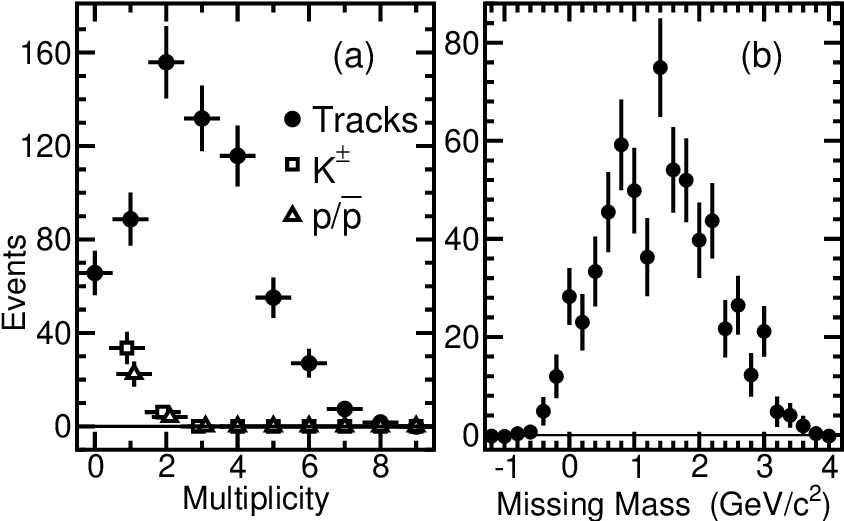}
\end{center}
\vspace{-0.6cm}
\caption {
  Background-subtracted distributions for the 649 \LcLcb events in 
  the data:
  (a) the numbers of additional tracks, identified $K^\pm$ and 
  identified \ppbar; and 
  (b) missing mass, with imaginary masses given negative real values.
  Most events have no identified $K^\pm$ or \ppbar and the
  corresponding zero-multiplicity points are off the vertical scale in (a).
  }
\label{fig:oparts}
\end{figure}

We investigate the structure of the \LcLcb events using the 
\Lc and \Lcb candidates
along with additional charged tracks that have at least 10 points 
measured in the DCH, 
5 in the SVT, and extrapolate within 5~mm of the beam axis.
We subtract appropriately scaled distributions in the background regions 
from those in the signal region
to obtain distributions for \LcLcb events.
Figure~\ref{fig:oparts}a shows the distribution of the number of
additional tracks, 
as well as the numbers of identified $K^\pm$ and \ppbar among them.
Were each $c$~baryon compensated by a light antibaryon, 
  then---assuming that half the antibaryons have an antiproton
  in the final state and 
accounting for \ppbar detection efficiency---we would expect
45\% of these events to contain one 
identified \ppbar and another 20\% to contain both an identified $p$
and a \pbar;
we observe only 3.4\% and 0.6\%, respectively.
Figure~\ref{fig:oparts}b shows the distribution of missing mass,
calculated from the four-momenta of the initial $e^+$ and $e^-$, 
the reconstructed \Lc and \Lcb, and 
all additional tracks
interpreted as pions.
A typical $\bary_{c1}\overline{n} X n\aary_{c2}$ event,
containing both a neutron and an antineutron,
would have a missing mass well in excess of 2~\gevcc.

The distributions in Fig.~\ref{fig:oparts}
indicate that the majority of the \LcLcb events do
not contain additional baryons, and therefore that the conservation of
baryon number is realized with the primary $c$ and \cbar hadrons.
  In the background-subtracted sample of $649 \pm 35$ \LcLcb signal events,
  there are 
  $28 \pm 6$
additional identified \ppbar candidates.
  These \ppbar candidates include background from two main sources:
  interactions in the detector material and misidentified pions or kaons.
  We expect 5 protons from material interactions.
  We also expect about 12 pions or kaons misidentified as protons,
  based on the numbers and momenta of the observed additional 
  $\pi^\pm$ and $K^\pm$ tracks.
  In cross-checks these expectations are found to be consistent
  with the data within uncertainties: there are $8 \pm 4$ more
  identified $p$ than \pbar (with the excess attributed to material
  interations), and there are $7 \pm 3$ events seen with exactly one
  additional identified \ppbar and an event missing mass below
  750~\mevcc (inconsistent with a missing second baryon, and so
  attributed to a misidentified kaon or pion).
Subtracting the expected contributions
  from these two background sources,
correcting for efficiency, and
assuming equal $p$ and $n$ production rates,
we estimate that we observe $13 \pm 8$ true four-baryon events.
This is well below the rate of 100 to 250 four-baryon events
expected for uncorrelated production,
let alone the observed rate of $649 \pm 35$ events,
indicating that the four-baryon process is strongly suppressed
and that the
  primary production
process dominates.

\begin{figure}[t]
\begin{center} 
\includegraphics[width=\linewidth]{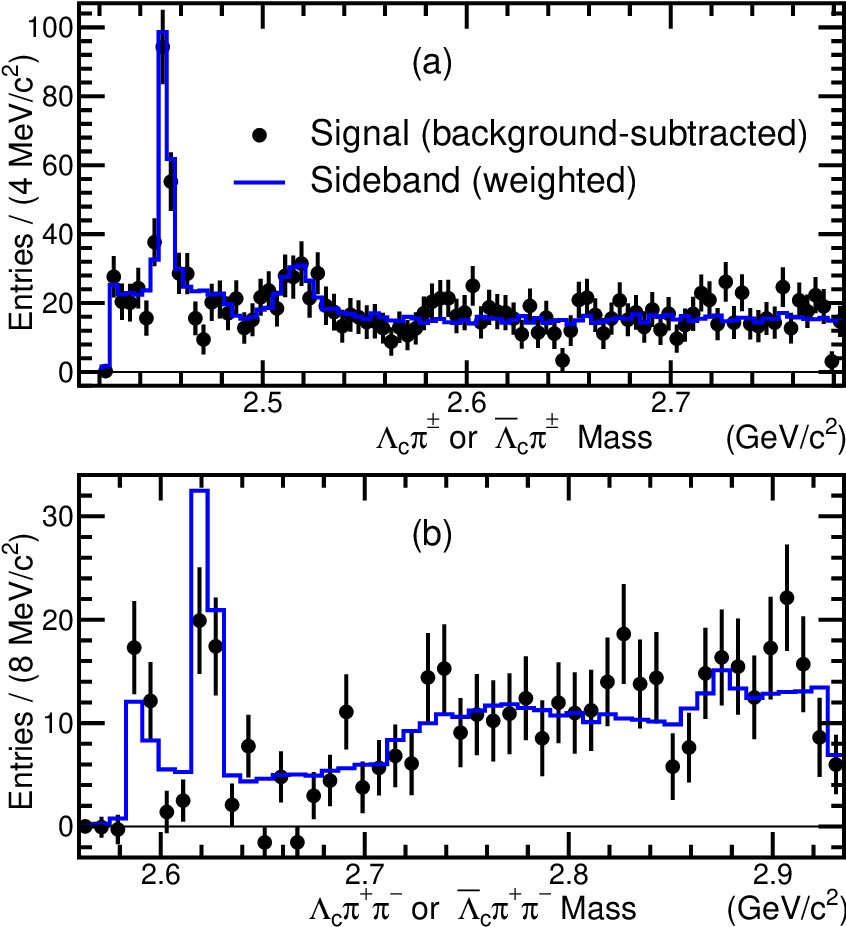}
\end{center}
\vspace{-0.7cm}
\caption {
  Invariant mass distributions for (a) $\Lc\pi^\pm$ and $\Lcb\pi^\pm$ and 
  (b) $\Lc/\Lcb\pip\pim$ combinations.
  The points with errors represent the background-subtracted \LcLcb
  events,
  and the weighted histograms are from the single-\Lc/\Lcb sidebands.
  }
\label{fig:lcpimass}
\end{figure}

None of the reconstructed events is consistent with
  the two-body process
$\epem \!\!\to\! \Lc\Lcb$.
However, the signal could arise from the
pair-production of $c$~baryons
if one or both are excited states that decay to \Lc/\Lcb:
$\epem \!\!\to\! \bary_{c1}\aary_{c2} \!\!\to\! \LcLcb$.
Combining \Lc/\Lcb candidates with one or two additional tracks
assigned the pion mass hypothesis gives the
invariant mass distributions in Fig.~\ref{fig:lcpimass}.
The points represent sideband-subtracted signal events and
the histograms the single-\Lc/\Lcb sidebands
with entries reweighted to reproduce the number
of the \Lc/\Lcb in signal events
  and their momentum and polar angle distributions in the lab frame.
Peaks are visible in the sideband data for the $\Sigma_c^{++/0}(2455)$,
$\Sigma_c^{++/0}(2520)$,
and the excited $\Lambda_c^+$ states at 2593, 2625, 2765 and 2880~\mevcc.
We find no unexpected peaks in our $\Lc\pi (\pi)$, $\Lc K$ or $\Lc\pbar$ 
mass distributions.
The points are consistent with the histograms,
indicating similar $c$~baryon compositions in the two event types.
Only two events are kinematically consistent with 
$\epem \!\!\to\! \bary_{c1}\aary_{c2}$ for these known $\bary_c$.
Distributions of $\theta$ and the decay angles in the $\Lc\pi$ rest
frames are consistent with multihadron events, 
and not with very heavy states decaying into a \Lc and more than two pions.
We conclude that 
$\epem \!\!\to\! \bary_{c1}\aary_{c2}$ 
processes
represent a small fraction of our sample.
From the fits in Fig.~\ref{fig:lcpimass},
we estimate that $35 \pm 3\%$ of the \Lc and
  $29 \pm 2\%$
of the additional pions in our sample
are decay products of heavier $c$~baryons.

Having established the presence of a category of events containing a
$c$~baryon, a \cbar baryon, no other (anti)baryons, and 
several intermediate mesons,
we study the number and structure of these mesons.
We exclude events with an identified \ppbar or 
a missing-mass-squared below $-$0.25~\gevcsq.
We estimate that the sample contains a further $5 \pm 5$
four-baryon events in which no \ppbar is detected; we
take these to have the same distributions as the events
with an identified \ppbar and subtract an appropriately
scaled contribution to correct for them.
In this sample of 
  $619 \pm 35$
events we study a number of quantities
including the \Lc/\Lcb and additional track momenta, polar angles,
rapidities and opening angles.
Their inclusive distributions
are quite similar to those in the single-\Lc/\Lcb sample 
and similar to those in all hadronic events.
In particular,
signing the thrust axis such that the \Lc rapidity is positive,
the \Lc and \Lcb rapidities cluster near $+$1.1 and $-$1.1 units,
respectively,
with the additional tracks of each charge distributed broadly and
symmetrically in between.

These 619 events contain only
  $45 \pm 10$
identified $K^\pm$
  of which about 20 are expected to be misidentified
  pions. The events
show no mass peak for \KS candidates
reconstructed from pairs of tracks
not included in the \Lc or \Lcb
(including tracks that do not extrapolate
within 5~mm of the beam axis).
The $K$:$\pi$ ratio is thus much lower than the value 0.3 typical of 
hadronic events, 
which might be due to the limited energy available 
and the fact that our $c$~baryons are non-strange 
(the lighter $c$-$s$ baryons do not decay into \Lc).
The $\pip\pim$, $K^\pm\pi^\mp$, and $\Kp\Km$ invariant mass distributions 
show no significant resonant 
  structure; in particular there is no evidence
for the $\rho^0$.
This implies a vector:pseudoscalar meson ratio much lower than the 
value near 1 typical of hadronic events,
and suggests that most tracks not from $c$~baryon decays
represent distinct intermediate mesons.

The intermediate meson multiplicity is distributed broadly.
We verify that the contribution from decays of heavier $c$~baryons
is not concentrated in any particular region in Fig.~\ref{fig:oparts}a,
but due to the limited sample size
we do not attempt to correct the distribution.
We observe an average of
  2.7
additional charged tracks per event. 
Correcting for $c$~baryon decays and tracking efficiency 
gives
  $2.6 \pm 0.2$
charged intermediate mesons per event,
where the uncertainty includes both statistical and systematic
effects.
The uncertainty is dominated by the track acceptance in these events, 
evaluated with a set of simulations based on the observed 
$\pi^\pm$ and $K^\pm$ distributions.
On average, the $c$ and \cbar baryons carry 75\% of the event energy, 
and the intermediate charged mesons account for
  about
65\% of the remainder.
This and the broad distribution of missing masses in
Fig.~\ref{fig:oparts}b suggest the presence of additional neutral mesons.
If intermediate $\pi^0$ are produced at half the $\pi^\pm$ rate,
as in typical hadronic events,
the average intermediate meson multiplicity would be~$3.9 \pm 0.3$.

The new type of event observed in our data might be explained by 
either primary diquark-antidiquark production 
or the production of multiple intermediate mesons between a baryon 
and antibaryon.
Neither the JETSET nor the HERWIG model produces events of the type observed, 
although both might be adapted to include one or both of the above 
processes.
JETSET does produce $\bary_{c1} M \aary_{c2}$ events,
where $M$ is a single meson,
often a vector decaying into two or three pions, 
but the event characteristics are far from consistent with the data.
Multiple intermediate meson processes occur naturally in the UCLA model, 
which also predicts an enhanced \LcLcb fraction due to events of this type, 
with suppressions of kaons and vector mesons.
The version of the UCLA model used
does not describe the observed events in detail, 
having an average of only 1.8 intermediate mesons with a distribution
peaked at low values, 
but the results presented here should encourage development of this 
and other relevant models.

In summary, 
we isolate a sample of
  $649 \pm 35$
\eeccb events containing
both a \Lc and a \Lcb
with high momentum in opposite hemispheres,
and study these events in detail.
The number of events is 
estimated to be
  about 4
times that expected 
if the leading $c$ and \cbar hadron types are uncorrelated,
confirming an observation by the CLEO Collaboration.
  Taking
  advantage of the 
  particle identification capabilities 
  of the \babar\ detector and the large data sample, we
  are further able to establish that
  almost all of these events contain
  no
additional baryons.
They do contain
  $2.6 \pm 0.2$
additional charged
intermediate mesons on average,
  and
events with zero additional mesons
do not contribute significantly.
  Our event sample exhibits
distributions of
momentum, angle, rapidity, and $c$~baryon type
similar to those in typical hadronic events,
but 
  contains
fewer kaons and vector mesons.
This is direct evidence for a new class of multihadron events,
in which 
baryon number is conserved by a leading baryon and antibaryon,
rather than locally along the hadronization chain.

We are grateful for the excellent luminosity and machine conditions
provided by our \pep2\ colleagues, 
and for the substantial dedicated effort from
the computing organizations that support \babar.
The collaborating institutions wish to thank 
SLAC for its support and kind hospitality. 
This work is supported by
DOE
and NSF (USA),
NSERC (Canada),
CEA and
CNRS-IN2P3
(France),
BMBF and DFG
(Germany),
INFN (Italy),
FOM (The Netherlands),
NFR (Norway),
MES (Russia),
MEC (Spain), and
STFC (United Kingdom). 
Individuals have received support from the
Marie Curie EIF (European Union) and
the A.~P.~Sloan Foundation.



\begin{thebibliography}{99}
\bibitem{params}
We use default parameter values for all models unless otherwise noted.
\end{thebibliography}
\end{document}